\documentclass{amsart}
\usepackage{graphicx}
\theoremstyle{definition}

\theoremstyle{remark}

\numberwithin{equation}{section}
\begin{document}

\title{ J-PARC P02 project:"THE Study of Exotic Multiquark States in
Systems with $\Lambda $-Hyperons, $K^0_S $-Mesons and $\gamma$ -
Quanta "}%
\author{P.Z. Aslanyan}%
\address{Joliot-Curie 6, Dubna, p.o. 141980, Moscow region, Russia.
Joint Institute for Nuclear Research}%
\email{aslanian@jinr.ru}%


% ----------------------------------------------------------------
\begin{abstract}
The designed 2m propane bubble chambers(PBC) with modern power
technologies for PC and high precision digital photographic methods
is a unique multi-propose, competitive capable and
higher-informative 4$\pi$ detector for study of exotic multi-strange
events with $V^0$($\Lambda,K^0_s$ and $\gamma$) particles, light
hyper-nucleus, ($V^0, V^0$) interactions and other correlations (P02
J-PARC LOI). The designed PBC will provide  on line data measurement
and data taking more than 100 times faster(or $>10^7$ events/year)
than old CERN-HPD system. First from all of unbeatable privilege for
PBC are registration of multi-vertex or complex decay modes(with
10-50$\mu$m space resolution), where is included of the beam area
too. The acceptance of beam area for detectors is crucial important
for $\Lambda$ hyperon physics, because more than 70\% from $\Lambda$
hyperons are emitted in the beam area with azimuth $\beta$ or polar
angles $< 15^0$ in p+C reaction at 10 GeV/c.

Strange multibaryon states with $\Lambda$- hyperon and $K_s^0$ meson
subsystems has been studied  by using data from 700000 stereo
photographs or $10^6$ inelastic interactions which was obtained from
expose 2-m propane bubble chamber LHEP, JINR to proton beams at 10
GeV/c.  The observed well-known resonances $\Sigma^{0}$,
$\Sigma^{*\pm}$(1385) and $K^{*\pm}$(892) from PDG are good tests of
this method. The subject of proposed P02 project  allow to explore
of multi-strangeness in hadronic systems with $V^0$ particles what
are also  included  as P05,P07,P18, p22 and p027 approved
experiments by a missing mass method at J-PARC. New developed
research complexes NUCLATRON-M at JINR and  J-PARC at JAEA(KEK) on
base of progress in digital technology for BC experimental method
allow to obtain of a necessary statistics in reasonable time, what
inspire the optimism for establish  exotic states with $\Lambda$ and
$K^0_s$ particles, which were observed more than 30 year historical
periods from different old experiments with a poor statistics. Only
with high statistics or from many similarly old photographs without
new necessary acceptance, resolution and methods analysis there will
not possible to obtain of new information about these objects.
\end{abstract}
\maketitle
% ----------------------------------------------------------------
\section{Preliminary list of  participants and institutes.
 }\label{part}

\begin{center}
\small \vspace*{1cm}  P.Zh.Aslanyan$^{1,\dag}$(contact
person\footnote{ $\dag$ {\it Email -
paslanian@jinr.ru}}),Y.Akaishi$^5$, T. Yamazaki$^4$,
P.Palazzi$^{10}$,Ch. Samanta$^{11}$, P. Banerjee$^{11}$,
A.A.Baldin$^1$, V.V. Burov$^3$,A. E. Dorokhov$^3$, N. D.
Dikoussar$^2$, V.V. Glagolev$^1$ V.N.Emelyanenko$^1$,
P.I.Zarubin$^1$,N.I. Kochelev$^3$, A.I. Malakhov$^1$, G.A.
Kozlov$^3$ V.I. Moroz$^1$,Y. Ohashi$^6$, V.S. Rikhkvitzkiy$^2$,
V.N.Penev$^{1,7}$, Stanislav Vokal$^{1,8}$, V.V.Uzhinskiii$^2$, A.
Kechechyan$^1$ M. Kadykov$^1$, S.Stecenko$^1$, A.Troyan$^1$,
A.Belyaev$^1$, A. Taratin$^1$,  G.A.Ososkov$^2$,  V.Baranov$^4$ P.
Evtoukhovitch$^4$, V. Kalinnikov$^4$, W. Kallies$^4$, N.
Kravchuk$^4$, A Moiseenko$^4$, V. Samoilov$^4$, Z. Tsamalaidze$^4$,
O. Zaimidoroga$^4$, V. Kirakosyan$^9$, A. Ayriyan$^2$, S.
Baginian$^2$
\end{center}

\vspace*{0.5cm} {\small
$^1$ Laboratory of High Energy Physics, JINR, Dubna, Russia\\
$^2$  Laboratory of Information Technology, Dubna, Russia\\
$^3$ Laboratory of Theoretical Physics, Dubna, Russia \\
$^4$ Laboratory Nucleus Problem, JINR, Dubna, Russia\\
$^4$ Department of Physics, University of Tokyo, Bunkyo-ku,
Tokyo 113-0033, Japan, and RIKEN Nishina Center, Wako, Saitama 351-0198, Japan \\
$^5$ KEK, High Energy Accelerator Research
Organization, Tsukuba, Ibaraki 305, Japan\\
 $^6$  Japan Synchrotron Radiation Research Institute, SPring-8\\
$^7$ Bulgarian Academy of Science, Sofia, Bulgaria \\
$^8$ University of P.J. Safarik, Kosice, Slovakia \\
$^9$ Yerevan State of University, Yerevan, Armenia \\
 $^{10}$ particlez.org,particle physics, phenomenology, CERN,
http://www.particlez.org/.\\
 $^{11}$ Saha Institute of Nuclear Research, Kolkata, India\\
 Welcome to collaboration. \\}
\section{ Scientific goals and scientific merits}\label{expbc}
\indent

\subsection{Scientific goals} \label{intro} \indent
 %\vskip 5mm

  Multi- quark states, glueballs and hybrids have been searched for
  experimentally for a very long time\cite{jaffe},\cite{kop}, but none is established.
   The PBC method allow to study of the following actual problems: in-medium
modification of strange hadrons, the origin of hadron masses, the
restoration of chiral symmetry, the confinement of quarks in
hadrons, the properties of cold dense baryonic matter and
nonperturbative QCD, $\Lambda$ yields and the structure of neutron
stars.

Exotic strange multibaryon states have been
 observed in the effective mass spectra of: $\Lambda \pi^{\pm}$,
  $\Lambda p$, $\Lambda p p$, $\Lambda p \pi $,$\Lambda \Lambda
$ and $\Lambda K^0_S$, $K^0_Sp$, $K^0_S\pi^{\pm}$ subsystems.The
experimental data  from the 2-m propane bubble chamber for
 systems with $\Lambda$ and $K^0_s$ states have been analyzed in
 detail in reports \cite{blois09}$-$\cite{ect10}.

Strange multi-strange clusters are an exiting possibility to explore
the properties of cold dense baryonic matter \cite{yamaz, gal}.
Although such states were predicted by Wycech earlier, but only
recently the availability of experimental facilities as E471(KEK),
FOPI(GSI), FINUDA(INFN), OBELIX(CERN) and DISTO has observed. In
particular for studying these kind of exotic nuclei, has delivered
first experimental results which triggered a vivid discussion and
project of  AMADEUS, INFN. Following \cite{rafel}, we assume that
the above experimental fact is due to the formation of a 'blob' of
QGP. A recent paper over viewing the status of the problem, from A.
Gal \cite{gal}: "It is clear that the issue of $\overline{K}^0$
nuclear states is far yet from being experimentally resolved and
more dedicated, systematic searches are necessary".

The problems on the interaction between hyperons and nucleons have
been poor studied until today. At the same time the study of the
processes of elastic dispersion hyperons+N yields much information
on the nature of hyperons and the character of hyperon-nucleonic,
hyperon - nucleus and hyperon-hyperon forces. Non-elastic processes,
such as $\Lambda+p\to \Sigma^+N$ are also interesting. The study of
YN scattering offers a unique way of probing the short range
behavior of Baryon-Baryon interactions due to the added degree of
freedom, the strangeness.

The study a Ericson fluctuations for $\Lambda$ hyperons, $\gamma$
quanta and protons momenta in backward direction, because the
character of deviations of experimental spectra from "smooth"
theoretical curve predicted within the framework of the fireball
model of cumulative effect \cite{ericson},\cite{matveev}.

At present the experimental situation is confused; so is theory. So
too is the complete absence of exotic mesons, and, except for the
recent discoveries, of exotic baryons as well.The classic example is
a baryon with positive strangeness, a $Z^+$ as it is known, with
valence quark content uudd$\overline{s}$.    The designed bubble
chamber as a new competitive technique will allow in the study of
multistrange states with $V^0$ particles.Because only with high
statistics or from many similarly old photographs without new
necessary acceptance, resolution and methods analysis there will not
possible to obtain of new information about these objects.

\subsection{Scientific merits}

\subsection{The propane bubble chamber}\label{pbc} \indent

PBC method  is the most suitable, higher-informative and
multi-propose 4$\pi$ detector(included beam range too)
 for study of  exotic multi-vertex states with $V^0$ particles \cite{pbc},\cite{pac10}.
 The average geometrical weights  for  $\Lambda$, $K^0_s$ and
 $\gamma$  are  to 1.34, 1.22 and 4.1 in p+propane collision at
  momentum 10 GeV/c, respectively.
The average effective mass resolution of ($V^0,V^0$),($V^0$,stoping
particles) system is equal to 0.5-1.0 \%. A  low beam intensity
(15-20)particles/spill can particularly compensate by using of large
chambers (as 2m PBC), large cross sections (p+propane 1450 mb, dead
time  is  to 5 sec,  5 events/spill), fast cyclic chambers,
secondary relativistic beams from $\Lambda$, $\Xi^-$ hyperons and
$K^+$, $K^-$, $K^0_l$-mesons.

The GEOFIT based on the Grind-CERN  program  is used  to measure the
kinematic  parameters of  tracks:  momentum(P), tg$\alpha$($\alpha$
- depth angle) and azimuthal angle($\beta$) from the  stereo
photographs. The momentum(P) of  resolution and  the average track
length (L)  for  charged particles are  to
 $\Delta$P/P=2\%,$< L>$ =12 cm for stopped  particles and $\Delta$P/P=10 \%,
 $<L>$ = 36 cm for nonstopped  particles.  The momentum  resolution  of
  $V^0$ from (1V-3C) fit  is  to $\Delta$P/P= 2%. The mean value of  error for
  the depth and azimuthal angles are to $\Delta$tg$\alpha$ = 0.0099 ($0.6^o$) $\pm$ 0.0002
   and  $\beta$ = 0.0052 ($0.3^o$)$\pm$ 0.0001 (rad.).

   The estimation of ionization, the peculiarities of the end track points of the
    stopped particles, allowed one to identify them.  Protons, $K^{\pm}$  and $\pi^{\pm}$ can
    exactly identified  by ionization  over the following momentum range: 0.13$< P_p<$0.9 GeV/c,
     0.05$<P_K< $0.6 GeV/c  and 0.025$<P_{\pi} <$ 0.3 GeV/c. Protons  can   separate from
     other particles  in the  momentum  range  of  $P_p <$ 0.9 GeV/c .

\begin{figure}[ht]
%\begin{center}
       {a)\includegraphics[width=70mm,height=50mm]{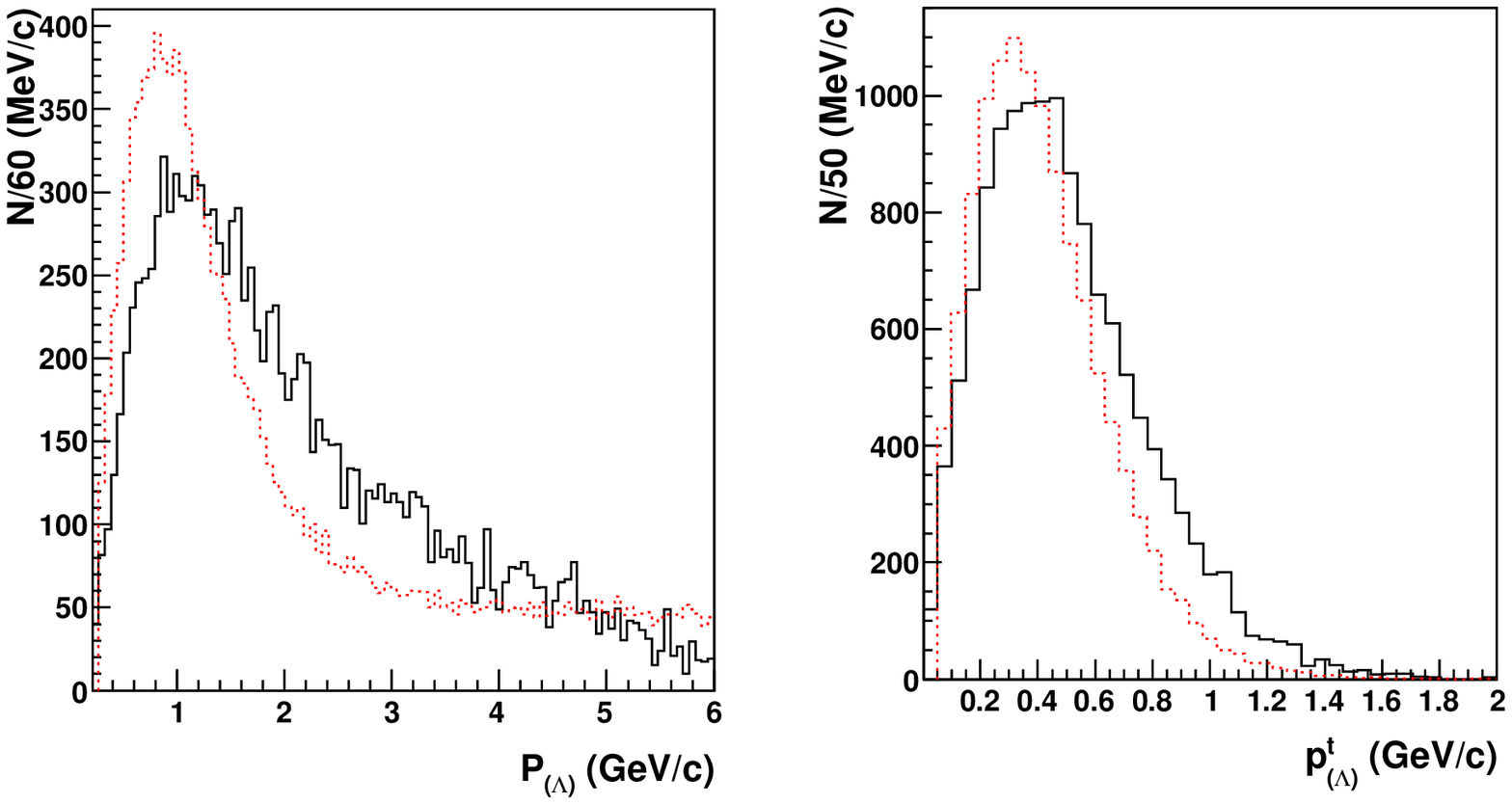}b)}
  {c)\includegraphics[width=70mm,height=50mm]{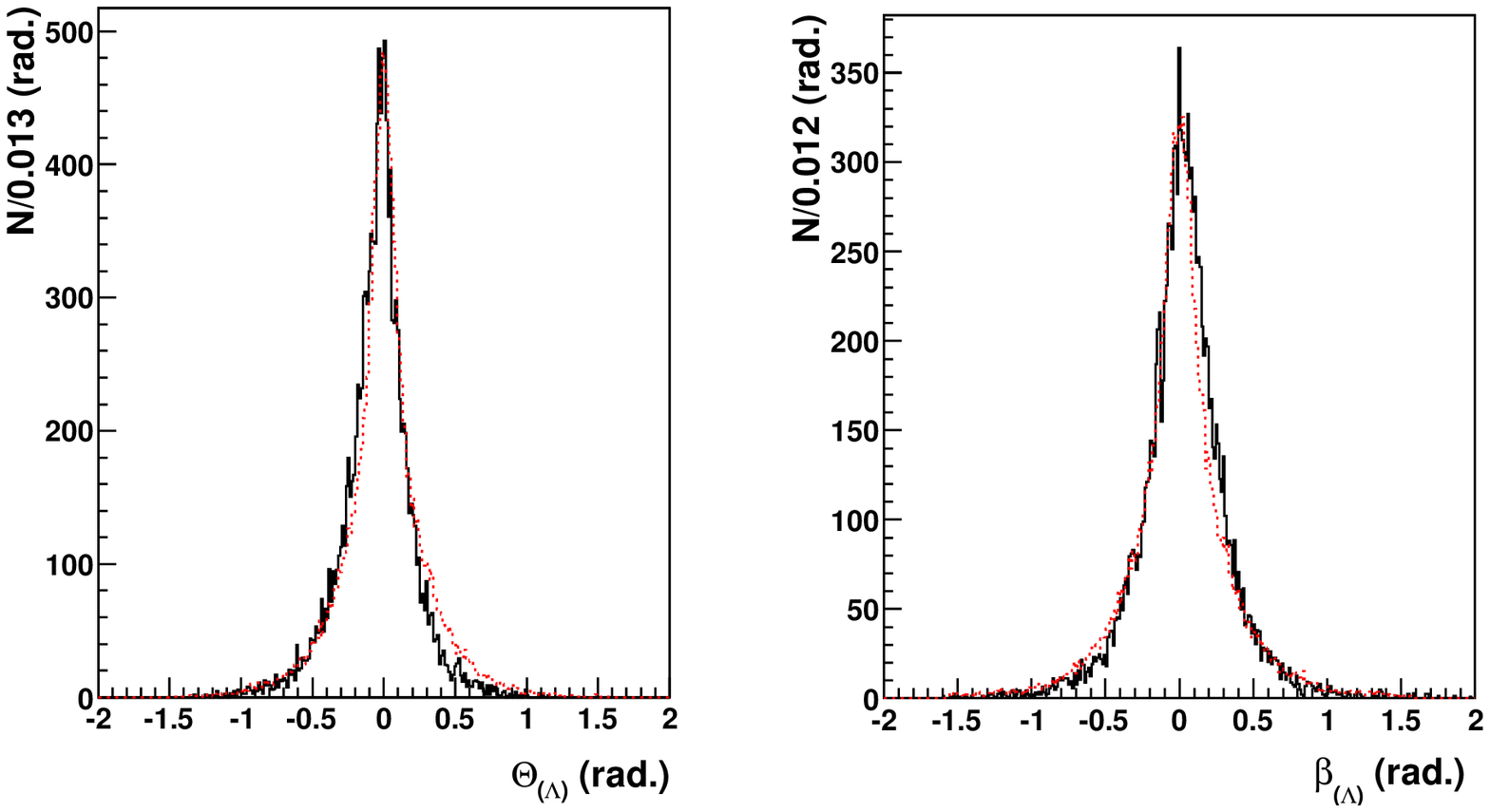}d)}
 \caption{Experimental(solid) and simulation by FRITIOF
model(dashed)  distributions of $\Lambda$ -  hyperons in p+propane
interaction at 10 GeV/c: a)by the momentum ($p_{\Lambda}$); b) by
the transverse momentum ($p^t_{\Lambda}$); c) by the polar angle
($\Theta_{\Lambda}$); d) by the azimuthal angle $\beta$. }
\label{lamptet}
%\end{center}
\end{figure}

 Figure~\ref{lamptet} shows basic experimental and simulation by FRITIOF model
  distributions of $\Lambda$ -  hyperons in p+propane interaction at 10 GeV/c. There
are satisfactory description by the polar angle ($\Theta_{\Lambda}$)
and d) by the azimuthal angle $\beta$. But FRITIOF model did not
describe of the momentum ($p_{\Lambda}$) and the transverse momentum
($p^t_{\Lambda}$) distributions. There are observed fluctuations by
momentum in  ranges of  1.56(4$\sigma$),1.9(3.5$\sigma$) and
2.15(3$\sigma$)GeV/c. Then observed fluctuations by azimuthal
$\beta$ angle (3$\sigma$)  in ranges of  -0.7$^o$  and  -6.8$^o$
(preliminary) has shown in Figure~\ref{lamptet},d. More than 70\%
from $\Lambda$ hyperons are emitted of beam area with azimuth or
polar  angles $<~15^o$ in p+C reaction at 10
GeV/c(Figure~\ref{lamptet}c,d).

\begin{figure}[ht]
    \begin{center}
        {a)\includegraphics[width=65mm,height=55mm]{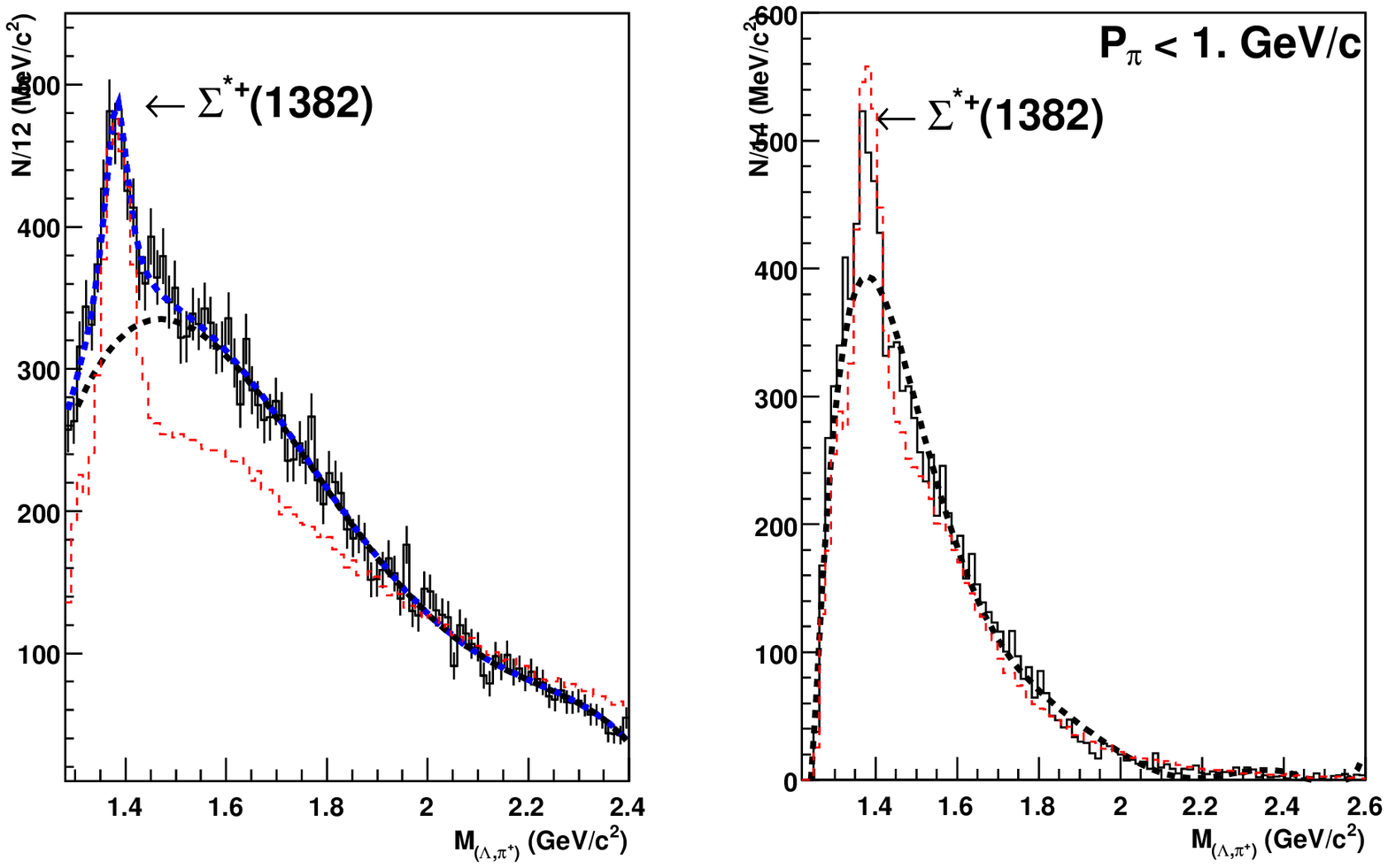}b)}
        {c)\includegraphics[width=65mm,height=55mm]{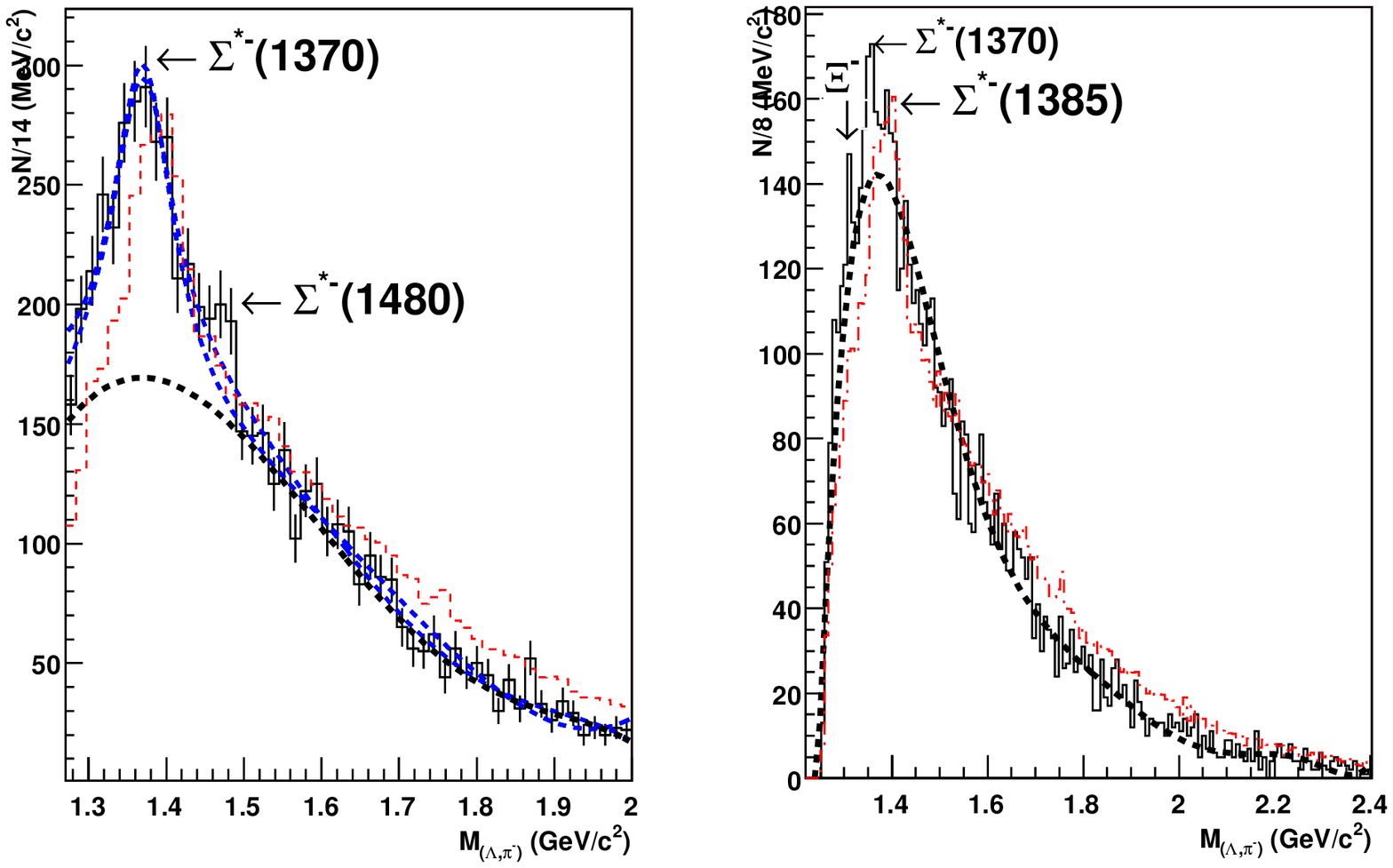}d)}
                 \caption{\it a)The $\Lambda \pi^+$ - spectrum for all
        combinations with
  bin size of 12 MeV/$c^2$
        b)  $\Lambda \pi^+$ - spectrum in momentum range of  $P_{\pi}<1$ GeV/c with
  bin size of 14 MeV/$c^2$;
        c) $\Lambda\pi^-$  spectrum for all combinations with
  bin size of 14 MeV/$c^2$;d)$\Lambda\pi^-$  spectrum for all combinations with
  bin size of 8 MeV/$c^2$. The simulated events by FRITIOF is the dashed histogram. The
background is the dashed curve.} \label{lpi}
  \end{center}
\end{figure}

The $\Lambda\pi^+$ effective  mass distribution for all 19534
combinations with bin size 12 MeV/$c^2$ has shown in Fig.\ref{lpi}a.
Fig.\ref{lpi}b  with cut of $p_{\pi^+}<1$GeV/c shown, where is
removed background from protons. This observed resonance identified
as $\Sigma^{*+}(1382) \to\Lambda \pi^+$  with similar decay
properties as in PDG data  which was a good test of this method.The
$\Lambda\pi^-$- effective mass distribution for all 6465
combinations with bin sizes of 14 and 8 MeV/$c^2$ in
Fig.\ref{lpi}(c,d) are shown. The width for $\Sigma^{*-}$ observed
$\approx$2 times larger  than PDG value(Fig.\ref{lpi},c). The peak
in mass range of M(1370) is  decayed into three
ranges:M(1317)+M(1360)+M(1385)(Fig.\ref{lpi},d). Where M(1317)and
M(1385) can interpreted as contributions from $\Xi^-$ and
$\Sigma^{*-}(1385)$. Then M(1360) peak can interpreted as
contribution from phase space or medium effect with
$\Sigma^{*-}(1385)$ in carbon nucleus.

\subsection{On line data taking and analysis from digital stereo
photographs}

This bubble chamber method was closed because there were two
defects. First defect is low beam intensity
(15-20)particles/spill(more 1 inter./sec ) which is particulary can
to compensate by using large length chambers as 2m PBC, large cross
sections($\sigma_{p+propane}\ge$1450 mb) , fast cyclic chambers,
parasitic and secondary exotic beams from $\Xi^-$, $\Lambda$
hyperons and $K^-,K^0_l$-mesons. Second defect is slow data
measurement and daq by using human eyes what now is removed  because
the development of digital photographic technology came up to higher
precision which is suitable for a bubble sizes($\approx$ 10-50
micron). Computers have evolved  from large mainframes to desktop
and laptops with computing power several orders  of magnitude larger
than what was available some  decades ago. The designed bubble
chamber detector with digital photographic stereo cameras by using
of software on base of CERN-HPD for reconstruction and automatic
analysis of stereo pictures what will provide  on line data taking
and measurements  more 100 times faster(or $10^7$ events/year) than
allowed  old HPD method. Automatic data taking software had been
developed for CHORUS, E373,E07 and OPERA experiments as a hybrid
emulsion methods.

 Main task of project will improve of the automatic on line data taking and
reconstruction software for bubble chamber methods on base of
obtained experience for digitized technologies from different
experiments in last years. A  estimation with 2 $\mu$m space
resolutions from  one projection of scanned stereo picture is equal
to $\approx$100 mb space in hard disc. Figure \ref{ming} shows the
"minimum guide" sketch of automatic on line and off line data taking
and measurement system flow chart from BC directly on base of HPD
CERN\cite{HPD}.

\subsection{DAQ  - PBC  experiment}

Beam momentum: 10 (or 12) GeV/c protons\\
Intensity as parasitic beam by using of bent crystal method: 15-20 protons/circle)\\
Spill interval: 5 sec/circle and 1 inter. per sec.\\
 p+$C_3H_8$ (inelastic interactions)/day:($>$) 72000\\
 (inelastic interactions from secondary particles)/day:($>$) 65000(2100 Tb)\\
Beam time: 100 day for setup\\
 Total number inelastic interaction from beam protons :7.2$\times 10^6$\\
 Target: $C_3H_8$ propane , 200cm length ,0.43 g/$cm^2$ density\\
 Estimated Yield: the number of identified $\Lambda (\to \pi^-p)$ hyperons $\ge$
 63000(40Tb) and $\Sigma^0(\to(\Lambda \gamma$)$\ge$ 2000\\
 Number of  $K^0_s(\to\pi^-\pi^+$) mesons $\ge$  30000(20 Tb) \\
 Statistical significance of reviewed peaks in
proposal p02  will  increase 4(5) times. \\

\subsection{The proposed stages of proposal}

\underline{Phase-1}: organize collaboration; select  main physics
tasks and reaction topology , inspect of old detector; discuss and
develop of type, size and cost for  universal super conducting
magnet; minimum guide program debug  on base of ROOT; manpower -7;
estimate of exactly budget for phase-1 and phase-2( $\approx$
370kUSD for PBC without magnet, 3 year), the budget for phase-1 is
37kUSD in  1 year.

\begin{figure}[ht]
{a)\includegraphics[width=70mm,height=60mm]{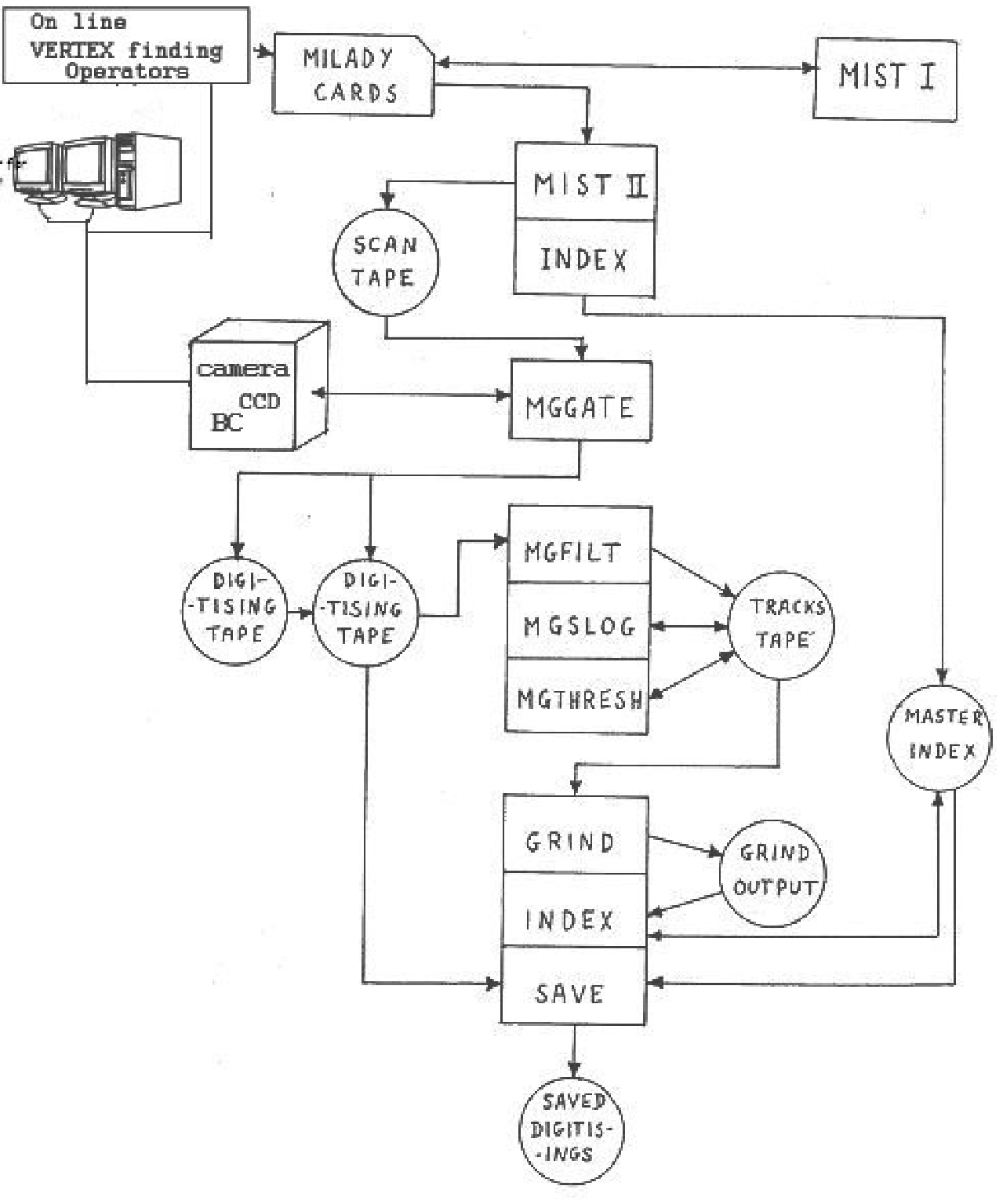}}
{b)\includegraphics[width=70mm,height=60mm]{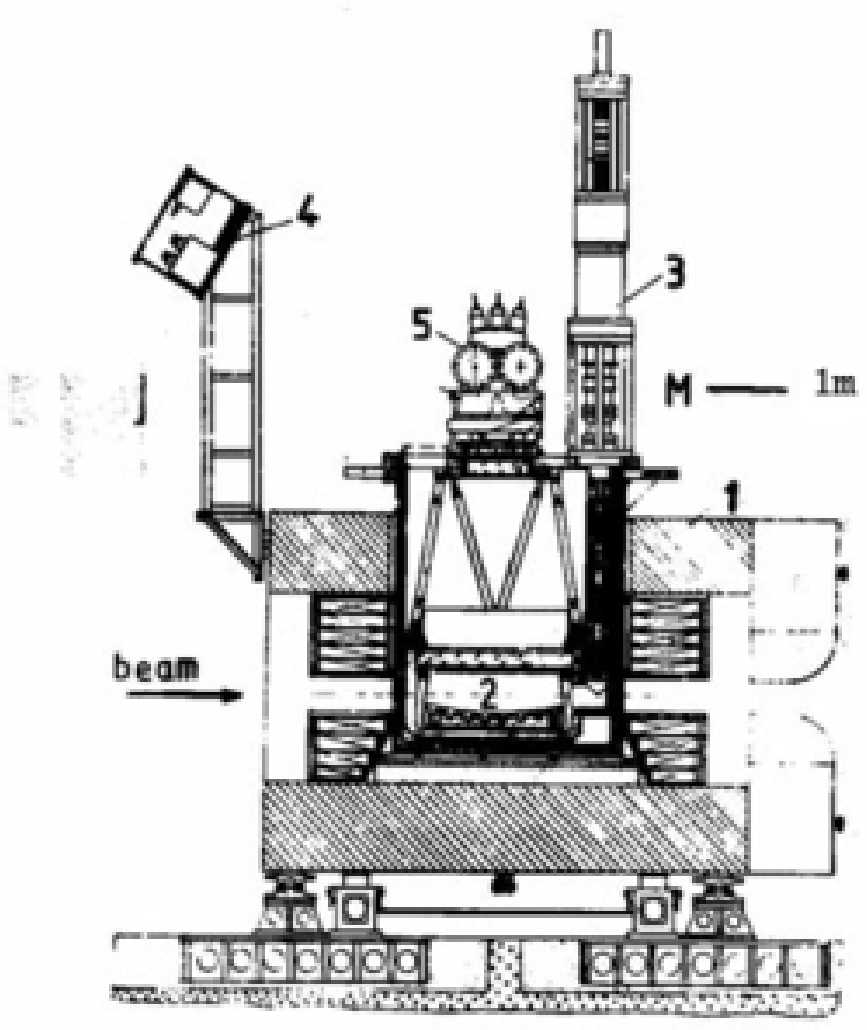}}
    \caption{a)The sketch of automatic on line
and off line data taking and measurement  system flow chart on base
of Minimum Guidance system for CERN HBC\cite{HPD}. b)The frontal
view of HBC,JINR, where 1 is a universal magnet as a barrel; 2 is
chamber;
 3 is mechanism of the expansion\cite{hbc}.} \label{ming}
\end{figure}

\underline{Phase-2}: design a proposed location , prepare CDR for
PBC without magnet(Fig. \ref{ming}) , prepare  CDR for parasitic
beam formation with bent crystal; develop and debug simulation and
reconstruction software on base of ROOT-GEANT4, install of minimum
guide  and "0" guide programs, construct of  PBC without magnet,
construct  and install of gas and other necessary systems, construct
and install of PBC with digital stereo photographic system;;  ,
manpower 13; CDR for universal super conducting magnet; prepare
total CDR for PBC with magnet;

\underline{Phase-3}: construct and install parasitic beam extraction
by bent crystal method, construct  of  universal magnet, construct
and install  of  PBC  with magnet,  construct and install trigger,
install data taking, test experiment for installation  of detector(3
year).

\bibliographystyle{amsplain}

\bibliography{}

\begin{table}
\begin{tabular}{|l|l|l|l|l|l|}  \hline
Stages&1 &2 &3&4&5  \\
&year&year&year& year&year \\ \hline phase
1&$\times$&$\times$&$\times$&-&-\\  \hline phase 2&-
&$\times$&$\times$&$\times$&$\times$\\ \hline
phase 3&-&$\times$&$\times$&$\times$&$\times$\\
\hline
\end{tabular}
\caption{Timetable for stages of proposal} \label{time}
\end{table}

\end{document}